\documentclass{jltp}
\usepackage{graphicx}
\usepackage{epsfig}

\title{Adsorption of {\it para}-hydrogen on krypton pre-plated graphite}
\author{Joseph Turnbull and Massimo Boninsegni} 
\address{Department of Physics, University of Alberta \\ Edmonton, 
    Alberta, Canada T6G 2J1}
\runninghead{J. Turnbull and M. Boninsegni}{Adsorption of {\it para}-hydrogen on krypton pre-plated graphite}

\def\he4{$^4$He}
\def\paraH2{{\it p}-H$_2$}
\def\Am2{\AA$^{-2}$}

\begin{document}

\maketitle

\begin{abstract}  
Adsorption of  \textit{para}-hydrogen on the surface of graphite pre-plated with a single
atomic layer of krypton, is studied theoretically by means of path integral ground state  
Monte Carlo simulations.  We compute energetics and  density profiles of {\it para}-hydrogen, and determine the structure of the adsorbed  film for various coverages.  Results show that there are two thermodynamically stable monolayer phases of \paraH2, both solid; one is commensurate with the krypton layer, and has coverage $ \theta $ $ \approx $ 0.0636 \Am2; the other is incommensurate, has coverage $\theta $ $ \approx $ 0.0716 \Am2, and is compressible up to
$\theta $ $ \approx $ 0.0769 \Am2.  No evidence is observed of a liquid phase at $T$=0 for intermediate coverages.  These results are qualitatively similar to what is seen for \paraH2 on bare graphite.  Quantum exchanges of hydrogen molecules are suppressed in this system.

\end{abstract}

\section{INTRODUCTION}

Low temperature adsorption of highly quantal fluids, such
as helium or {\it para}-hydrogen (\paraH2), on a variety of substrates,
has been the focus of much experimental and theoretical work.  A major
motivation of these studies is the exploration of the fascinating properties
that such adsorbed quantum films display, often considerably different
than those of the bulk materials.

In particular, a fluid of \paraH2 molecules is an interesting physical system for a number of 
reasons. Because a \paraH2 molecule has half the mass of a helium atom, 
zero-point motion can be expected to be quite significant; each molecule is
a spin-zero boson, and therefore it is conceivable that, at low enough temperature,
a \paraH2 fluid might display physical behavior similar to that of fluid 
helium, including superfluidity.\cite{ginzburg72}

Unlike helium, though, bulk {\it p}-H$_2$ solidifies at low temperature 
($T_{\rm c} \approx$ 14 K); this prevents the observation of phenomena such as 
Bose Einstein condensation (BEC) and, possibly, superfluidity (SF), which are speculated
to occur in the liquid phase below $T$ $\approx$ 6 K.  Solidification is due the depth of the 
attractive well of the potential between two hydrogen molecules, significantly greater than that between two helium atoms. Several, attempts have been made\cite{bretz81,maris86,maris87,schindler96} to supercool bulk liquid \paraH2, but the search for SF (in the bulk) has so far not met with success.

Potential avenues to explore, toward stabilizing a liquid phase of \paraH2 to low enough temperatures that a SF transition could be observed, include the reduction of dimensionality.  An extensive theoretical study of the phase diagram of \paraH2 in two dimensions (2D) has been recently carried out, based on Path Integral Monte Carlo (PIMC)  simulations.\cite{boninsegni04b} The main result is that the equilibrium phase of the system at low $T$ is a triangular crystal, with a melting temperature $T_{\rm m}\sim$ 6.8 K. This value is approximately half that of bulk \paraH2, but  still significantly higher than the temperature at which the system, if it remained a liquid, would undergo Bose Condensation and turn superfluid, estimated at $\sim$ 2 K in 2D. Another important result of the same study, is that no {\it metastable} liquid phase exists; for, the system remains a liquid down to the spinodal density, below which it breaks down into solid clusters.  These results, in part,  cast  doubts on the prospects of observing a liquid (SF) phase of \paraH2 at low $T$, in 2D. 

The closest experimental realization of a two-dimensional system, is a film of \paraH2 molecules adsorbed on a substrate. However, quantum zero-point motion of adsorbed particles in the direction perpendicular to the substrate can be quite significant, as calculations for adsorbed helium films on alkali metal substrates show.\cite{boninsegni99} One might speculate that such zero-point motion may result in an effective screening of the \paraH2 intermolecular interactions, possibly leading to a reduction of $T_{\rm m}$, with respect to the purely two-dimensional case. Indeed, some PIMC studies\cite{boninsegni04_li} of \paraH2 films adsorbed on a lithium substrate yielded a melting temperature of approximately 6.5 K, i.e., slightly lower than in 2D. The interesting physical question is whether $T_{\rm m}$ may be reduced even further, by an appropriate choice of substrate, to the point where collective quantum many-body phenomena could become observable, in the liquid phase.

Experimentally, adsorbed films of \paraH2 on surfaces are readily accessible, and indeed have been studied extensively on a variety of substrates.  For example, various techniques\cite{nielsen80,lauter90,wiechert91,vilches92} have been used to study the phase diagram and structure of monolayer \paraH2 films adsorbed on graphite.  In a recent neutron scattering investigation of deuterium (D$_2$) films adsorbed on a krypton pre-plated graphite substrate,\cite{wiechert04} evidence of a stable liquid phase of  D$_2$ down to $T\sim$1.5 K was reported.  Such a result is truly remarkable, particularly considering that it pertains to the heavier isotope of hydrogen, which should display an even stronger tendency to crystallize than \paraH2. Motivated by this finding, we have undertaken a theoretical study of the low temperature phase diagram of \paraH2 films adsorbed on krypton pre-plated graphite. 

On general grounds, one would expect \paraH2 to remain liquid as well, on account of its lighter mass, even though the intermolecular interaction is slightly more attractive for \paraH2.\cite{wang03,shi03}  We explore a fairly simple model of our  system of interest in its ground state ($T$=0 limit) to serve as a baseline for future, more refined calculations.  Energetic and structural properties of a layer of \paraH2 molecules adsorbed on Kr/Graphite are investigated 
theoretically, by means of Path Integral ground state (PIGS) Monte Carlo  simulations.  

The main results of this study are the following:
\begin{enumerate}
\item {A stable commensurate solid \paraH2 monolayer exists, with coverage (i.e., 2D density) $\theta_\circ$=0.0636 \Am2.  This solid is commensurate with the krypton plating.}
\item {An incommensurate solid \paraH2 monolayer exists, with coverage $\theta_{1}$=0.0716 \Am2, compressible up to $\theta_{2}$=0.0769 \Am2.}
\item{No evidence is observed of a  thermodynamically stable {\it liquid} phase in the ground state of this system.}
\item{Evidence of quantum exchange between \paraH2 molecules is not found in the $T$=0 limit; this in turn indicates absence of superfluid behavior in this system.}
\end{enumerate}

The remainder of this manuscript is organized as follows:  Sec. \ref{model} offers a description of the model used for our system
of interest, including a discussion of the potentials and  justifications for the main underlying 
assumptions.  Sec. \ref{method} involves a brief discussion of the computational technique and specific details of its implementation, in addition to details of calibration and optimization.  The results are presented in Sec. \ref{results}; finally, Sec. \ref{conclusions} is a summary of the 
findings and our concluding remarks.

\section{MODEL} 
\label{model}

We consider a system of $N$ \paraH2 
molecules, sitting above a substrate consisting of a single atomic layer of krypton, below which is a graphite substrate, which we assume to be smooth. 
The ($L$) krypton atoms are assumed fixed in space at positions ${\bf R}_k$, owing to their relatively large mass.  They sit at a height of 3.46 \AA\ over the graphite; this distance corresponds to the minimum of the most accurate Kr-graphite interaction potential currently available.\cite{steele,gooding} All of the Kr atoms and \paraH2 molecules are regarded as point particles.  The model quantum many-body Hamiltonian is therefore as follows: 
\begin{eqnarray}\label{hm}
\hat{H}=-\frac{\hbar^{2}}{2m}\sum_{i=1}^{N}\nabla_{i}^{2}  +
\sum_{i<j}V(r_{ij})  +
\sum_{i=1}^N\sum_{k=1}^{L} U (|{\bf r}_i-{\bf R}_k|)  +
\sum_{i=1}^N {\tilde U}({z}_i) 
\end{eqnarray}

Here, $m$ is the mass of a \paraH2 molecule, 
$\{{\bf r}_j\}$ 
(with $j$=1,2,...,$N$) are the positions of the \paraH2 molecules, $r_{ij}\equiv |{\bf r}_i-{\bf r}_j|$; $z_{i}$ is the height of the $i$th \paraH2 molecule above the graphite surface. 
$V$ is the potential describing the interaction between any two \paraH2 molecules, and
$U$ represents the interaction of a \paraH2 molecule with a Kr atom.  

As mentioned above, the underlying graphite substrate is regarded as smooth. The justification for this assumption is that, due to the presence of the Kr spacer layer, the \paraH2 molecules  remain at a distance of at least $\sim$ 7 \AA\ from the graphite substrate. Thus, the effect of the corrugation of the substrate should be  negligible. Therefore, we use a simple ``3-9" potential to describe the interaction of \paraH2 molecules with the smooth graphite substrate.\cite{gatica04}  $\tilde{U}$ has the following form: 
\begin{eqnarray}\label{sim}
\tilde U(z_{i}) = \frac{4 C^{3}}{27 D^{2} z^{9}} - \frac{C}{z^{3}}
\end{eqnarray}
where $C$=7913.24 \AA$^{3}$ K and $D$=259.39 K are parameters derived from the original \paraH2-C Lennard-Jones parameters\cite{levesque02} ($\sigma=3.18$ \AA, $\epsilon=32.05$ K) and the density of carbon atoms in graphite ($\rho=0.114$ \AA$^{-3}$).  

All pair potentials are assumed to depend only on relative distances.  
The interaction  $V$ is described by the Silvera-Goldman potential,\cite{Silvera1} which provides an accurate description of energetic and 
structural properties of condensed \paraH2 at ordinary conditions of 
temperature and pressure.\cite{johnson96,operetto}   

The interaction of a \paraH2 molecule
 and a Kr atom is modeled using a 
standard 6-12 Lennard-Jones (LJ) potential; to our knowledge, there are no published values for the parameters of this potential for this specific interaction. Therefore, we make use of the Lorentz-Berthelot mixing rule,\cite{wang97,macgowan86} yielding $\epsilon=75.6$ K and 
$\sigma=3.3$ \AA.

The model (\ref{hm}) clearly contains important physical simplifications, such
as the neglect of zero-point motion of Kr atoms, 
as well as 
the restrictions to additive pairwise interactions (to the exclusion of,
for example, three-body terms), all taken to be central, and the use of the 
highly simplified LJ and "3-9" potentials. Nonetheless, it seems a reasonable starting point, and even quantitatively we expect it to capture the bulk of the physical picture.

\section{COMPUTATIONAL METHOD}  
\label{method}

Accurate ground state expectation values for quantum many-body systems
described by a Hamiltonian such as (\ref{hm}) can be computed 
by means of Quantum MOnte Carlo (QMC) simulations. 
In this work, the method utilized is \textit{Path Integral ground state} 
(PIGS), which is an extension to zero temperature of the standard, 
Path Integral Monte Carlo method.\cite{Ceperley1}  
PIGS    is a projection technique, which filters the exact ground state wave 
function out of an initial trial state. It is therefore closely related to
other ground state projection methods, such as Diffusion Monte Carlo (DMC),
but has a few distinct advantages (for a discussion, see, for instance, Ref. 
\onlinecite{Sarsa1}).  
Because this method is described in a number of publications, it will not be reviewed here. Some of the technical details of the calculation performed in this work (mainly, the short imaginary time propagator) are the same as in Ref. \onlinecite{boninsegni04a}.

The trial wave function utilized is of the Jastrow type:
\begin{equation}\label{trial}
\Psi_T({\bf r}_1,{\bf r}_2,...{\bf r}_N)= \biggl ( \prod_{i<j}^N e^{-v(r_{ij})}\biggr ) \times \biggl (\prod_{i=1}^N\prod_{k=1}^L 
e^{-u(|{\bf r}_i-{\bf R}_k|)}\biggr ) \times \biggl ( \prod_{i=1}^N e^{-w(z_i)}\biggr )
\end{equation}
 with
pseudo-potentials $w$ (\paraH2-graphite), $u$ (\paraH2-Kr), and $v$ (\paraH2-\paraH2) chosen as follows:
\begin{equation}
w(r)=\frac{\alpha}{z^3} \;\; , \;\; u(r)=\frac{\gamma}{r^5} \;\; {\rm and}  \;\; v(r)=\frac{\mu}{r^5}
\end{equation}

The values of the parameters $\alpha=30$ \AA$^{3}$, $\gamma=250$ 
\AA$^{5}$ and $\mu=750$ \AA$^{5}$  were obtained empirically, by minimizing the energy 
expectation value computed in separate variational calculations.  Using the trial wavefunction as defined above, 
we observe convergence of the ground state energy 
estimates with a projection time 0.250 K$^{-1}$, using a time step $\tau = 7.8125\times 10^{-4}$ K$^{-1}$.  

PIGS calculations for a range of \paraH2 coverages were carried out, starting from an initial configuration of 
para-hydrogen molecules sitting atop the Kr layer.  The simulation cell consists of a 6$\times$8  triangular lattice of
Kr atoms with 4.26 \AA\ nearest neighbor spacing. Periodic boundary conditions are used in the three directions, but the simulation cell is chosen sufficiently big in the $z$ direction that  they have no effect vertically.  Because of the strongly attractive character of the composite substrate, for a small 
enough number of hydrogen molecules, the system remains vertically
confined (i.e., \paraH2 molecules do not evaporate). 

The systematic errors of our calculation are attributable to finite
projection time  and the finite time step $\tau$.  
Based on comparisons of results obtained from simulations with different 
values of projection time and time step, we estimate our combined systematic error on the 
total energy per \paraH2 molecule to be of the order of 0.7 K or less (below 0.5\%).  

\section{RESULTS}  
\label{results}

Physical quantities of interest include the ground state energy per \paraH2
molecule, $e(N)$, and the vertical \paraH2 density profile, $\rho(z)$, above the composite substrate. These quantities can be computed in an unbiased fashion, i.e., the variational bias arising from the initial state (i.e.,  trial wave function (\ref{trial})) can be essentially completely removed.\cite{Sarsa1}
 
\begin{figure}
\centerline{\includegraphics[height=2.75in]{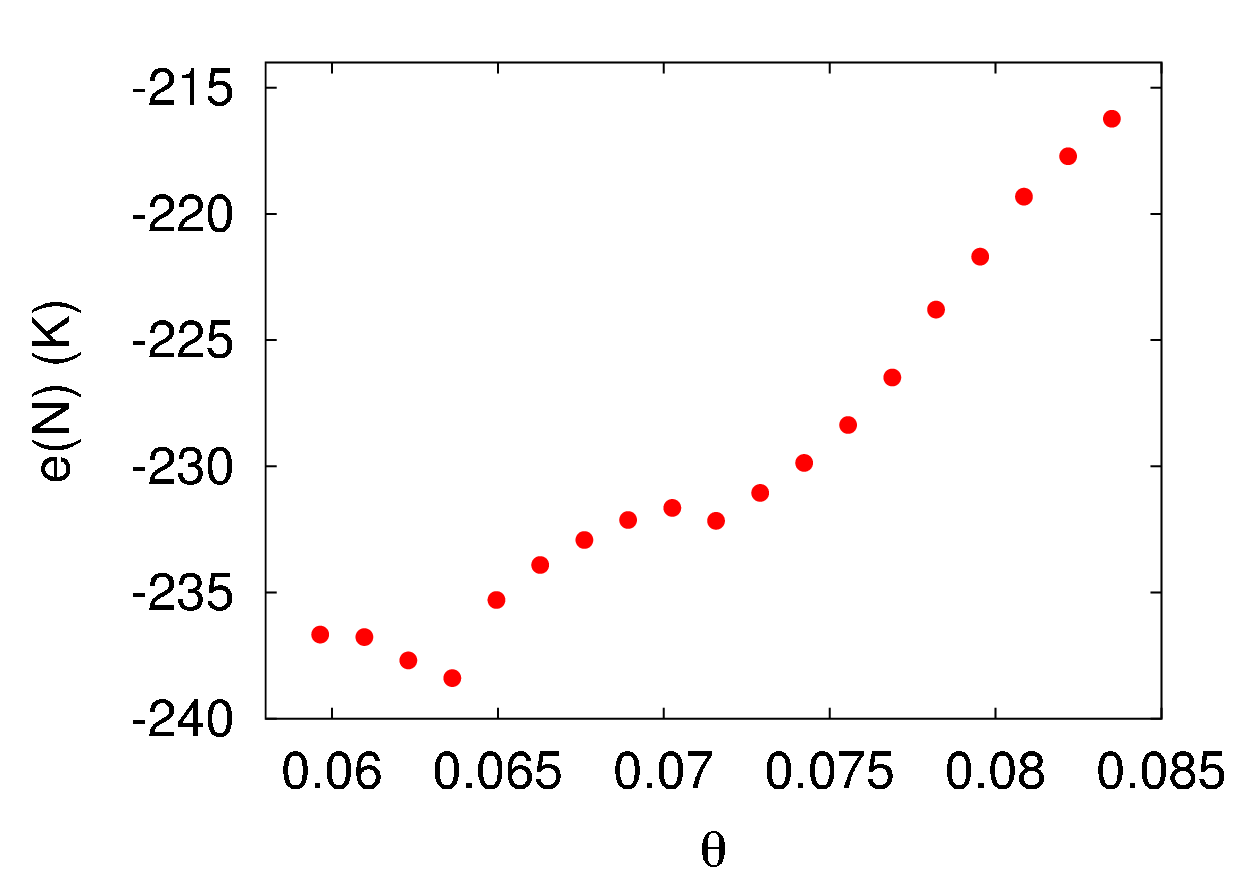}}
\caption{Energy per \paraH2 molecule $e(N)$ (in K) computed by PIGS, as a function of the coverage $\theta$ (in \AA$^{-2}$).}  
\label{energyplot1}
\end{figure}

Results for $e(N)$ are shown in Fig. \ref{energyplot1}.  The main features are two energetic minima, one at $\theta_\circ = 0.0636$ \AA$^{-2}$, a coverage corresponding to commensuration ($N$=48 \paraH2 molecules). The second, at $\theta_{1} = 0.0716$ \AA$^{-2}$, corresponds to an incommensurate solid monolayer, which remains stable, based on an analysis of the associated chemical potential (see below), up to $\theta_{2} = 0.0769$ \AA$^{-2}$ ($N$=58 \paraH2 molecules).  

A direct comparison can be made between our $e(N)$ curve, and that obtained by Nho and Manousakis (Ref. \onlinecite{nho03}), who studied \paraH2 monolayer films on bare graphite (at low temperature).  The shape of the energy curve in both cases is very similar; there is an energetic minimum at commensuration, corresponding to precisely the same coverage, followed by a negative-curvature increase.  Where we find a second energetic minimum, they find a change in curvature, in both cases preceding a positive curvature (thermodynamically stable) portion, although, on Kr pre-plated graphite, such a region seems more extended.  

\begin{figure}
\centerline{\includegraphics[height=2.75in]{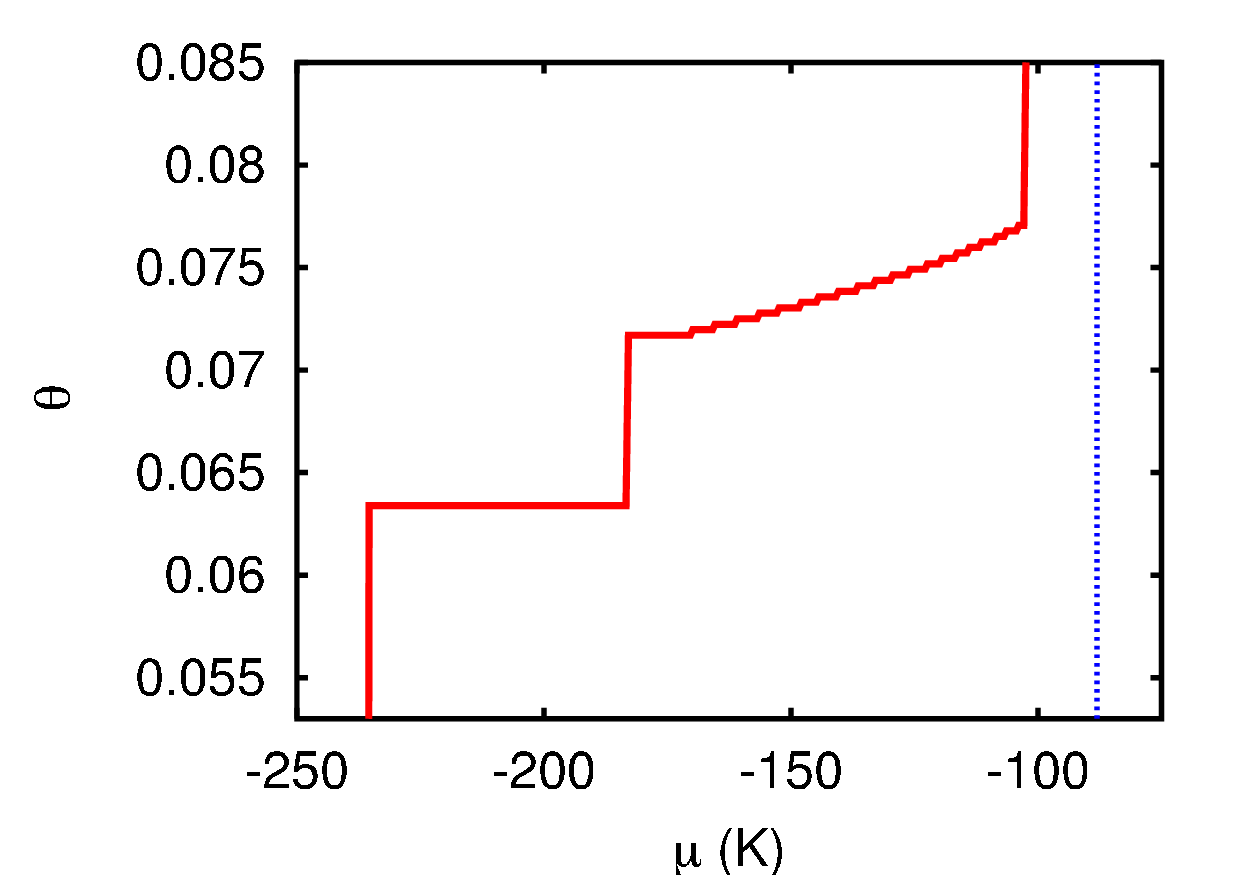}}
\caption{Coverage of \paraH2 molecules, $\theta$, ( \AA$^{-2}$) as a function of the chemical potential, $\mu$  (in K), computed as explained in the text.  Dotted vertical line shows the chemical potential of bulk solid \paraH2, at the $T$=0 equilibrium density (from Ref. \onlinecite{operetto}).}  
\label{chempot}
\end{figure}

We calculate the chemical potential, shown in Fig. \ref{chempot}, by first fitting the curve for $e(N)$ and then minimizing the grand canonical energy $\phi(N)=N(e(N)-\mu)$ with respect to $N$, for different values of $\mu$.  As shown by the data in Fig. \ref{chempot},  these are the only stable coverages, at least up to $\theta \approx 0.0848$ \AA$^{-2}$ (the highest coverage explored in this work).  The presumption is that the next thermodynamically stable configuration would be at second layer completion, $\theta \approx 0.127$ \AA$^{-2}$.

\begin{figure}[t]
\centerline{\includegraphics[height=2.65in]{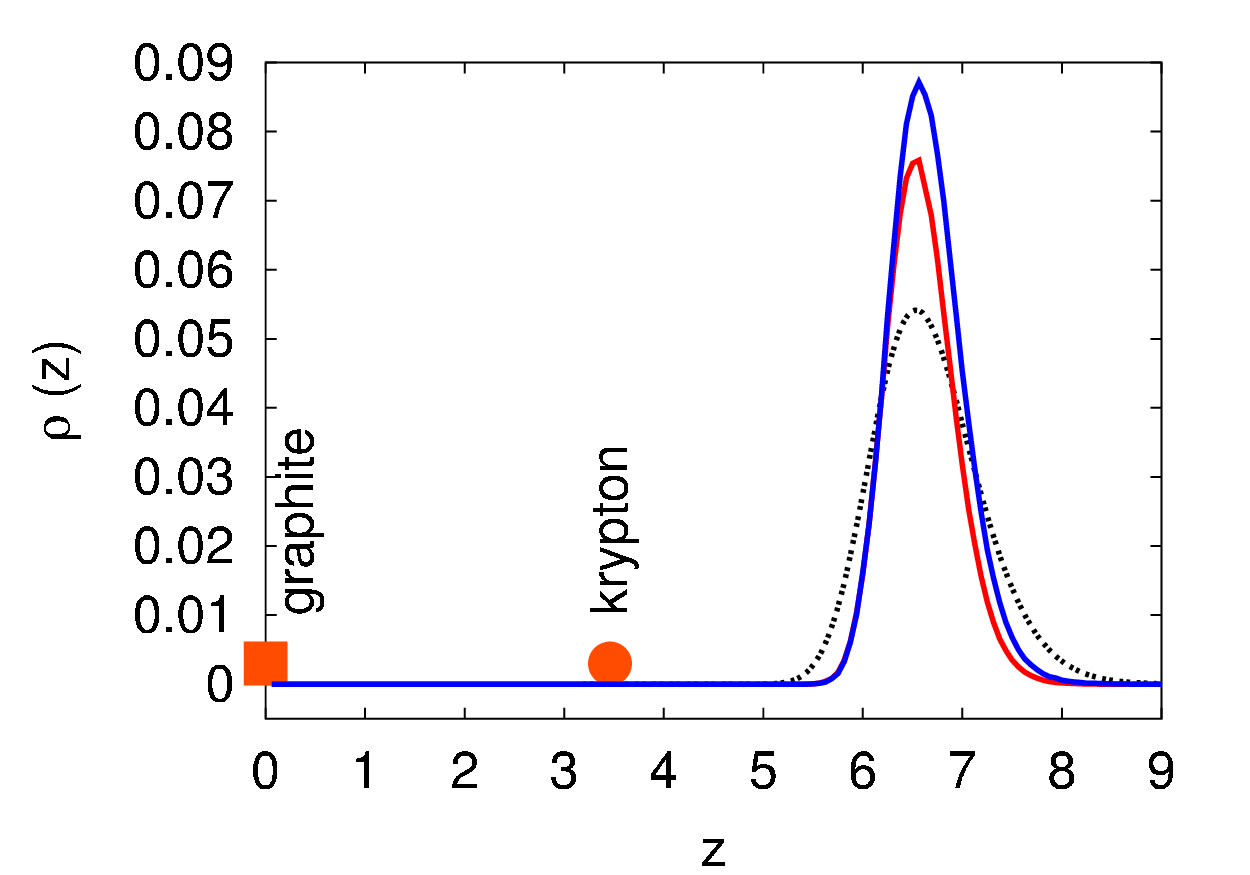}}
\caption{Density profile $\rho(z)$ (in \AA$^{-3}$)  of adsorbed \paraH2 at a coverage $\theta_\circ$  (solid line with lower maximum value), and at  $\theta_{2}$  (solid line with higher maximum value). The square represents the position of the graphite substrate and the circle that of the Kr layer.  Dashed line is the density profile for \paraH2 on a lithium substrate at the coverage $\theta$=0.070 \Am2 (from Ref. \onlinecite{boninsegni04_li}; it has been shifted along $z$ for direct comparison).  Distances are expressed in \AA.}   
\label{profile}
\end{figure}

Fig. \ref{profile} shows the (three-dimensional) \paraH2 density profile $\rho(z)$ for a coverage $\theta_\circ$ (commensurate solid layer), as well as the \paraH2 density profile for the highest stable coverage, $\theta_{2}$ (incommensurate solid layer).  
Also shown in Fig. \ref{profile} is the density profile for \paraH2 on lithium\cite{boninsegni04_li} at low temperature (2 K).  It is evident that for the substrate studied here, \paraH2 is more localized in the direction perpendicular to the substrate (the density profiles are peaked much more strongly) in comparison, despite being at $T$=0.   Thus, the physics of \paraH2 is considerably more 2D on this substrate, than on the weak Li substrate.  This is consistent with the stronger substrate attraction, and leads us to predict a melting temperature for the adsorbed film of the order of $\sim$ 7 K, i.e., comparable to that of purely 2D \paraH2. Obviously, however, only finite temperature calculation can address this point quantitatively. 
We also see that the profile for $\theta_{2}$ is extended slightly further to the right than that for $\theta_\circ$; in accommodating more \paraH2 in the first layer, the molecules are ``squeezed'' to occupy regions higher above the substrate.  

\begin{figure}
\centerline{\includegraphics[height=3in]{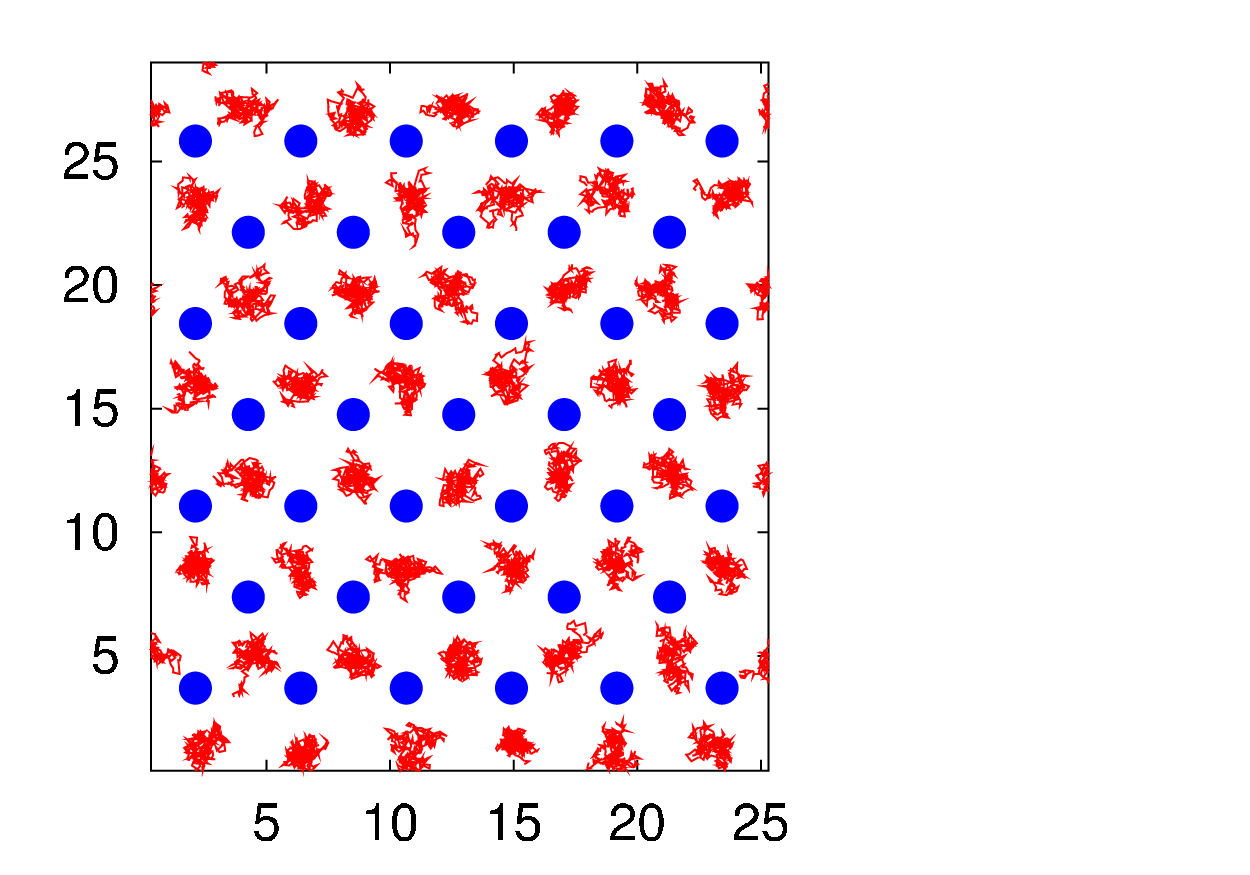}}
\caption{Snapshot of a typical configuration of \paraH2 molecules adsorbed to the Kr/graphite substrate at a coverage of $\theta_\circ$=0.0636 \AA$^{-2}$.  Large circles are Kr atoms.  The positions of all molecules at each one of 320 imaginary time slices are shown as discrete paths. Distances are expressed in \AA.}  
\label{snap1}
\end{figure}

\begin{figure}
\centerline{\includegraphics[height=3in]{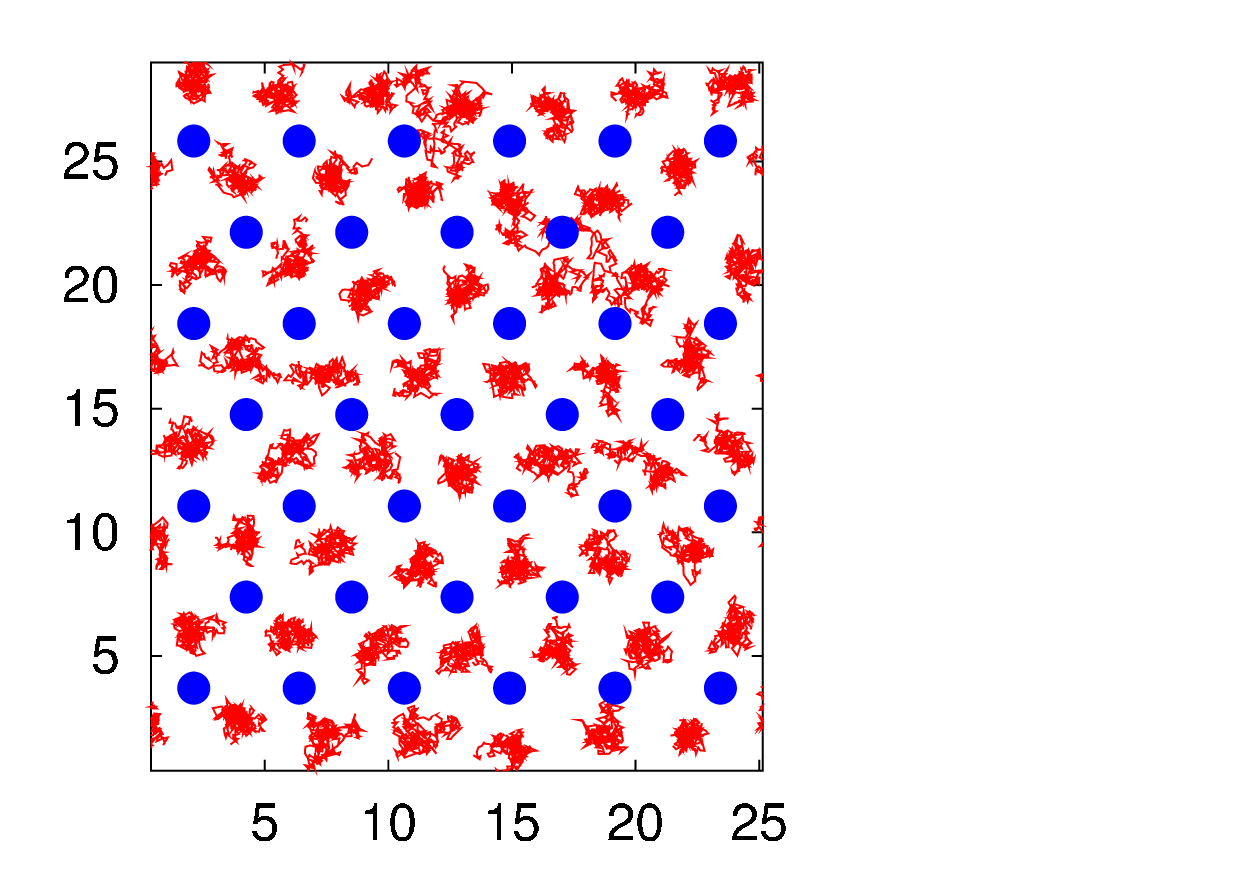}}
\caption{Same as Fig. \ref{snap1} but at a coverage of $\theta_{2}$=0.0769 \AA$^{-2}$}
\label{snap2}
\end{figure}

Snapshots of typical configurations of the \paraH2 solid film are displayed in Figs. \ref{snap1} and \ref{snap2}. The first, at $\theta_\circ$, shows the commensurate structure of \paraH2 corresponding to this system's first energetic minimum.  The second is a snapshot at the highest stable coverage, $\theta_{2}$; \paraH2 form an incommensurate solid, not found to be rotated relative to the Kr lattice (although the size of our simulated system clearly limits our capability of resolving such an issue).  56 of the 58 molecules in the simulations appear to try to form a 7$\times$8 regular triangular lattice, with the two additional molecules packing in, giving rise to the clear dislocations.  Simulations with higher coverages (thermodynamically unstable) were consistently found to put additional particles in the second layer.

In contrast, for \paraH2 on bare graphite,\cite{nho03} the somewhat stronger adsorption potential allows for an even denser packing; compression of the incommensurate phase is reported well beyond $\theta=0.0849$ \Am2, with the \paraH2 lattice rotated 5$^o$ relative to the graphite lattice.  We should also mention that, due to the system size and geometries employed in this work, we could not observe such crystal phases as the uniaxially compressed and the stripe one, which have been experimentally observed and theoretically studied for \paraH2  on graphite;\cite{nho03} we presume that these phases should exist in this system as well, and can certainly be studied with the computational method used here. However,  based on the results of Ref. \onlinecite{nho03} we do not think that their inclusion would significantly alter our main conclusion, concerning the existence of a liquid phase, which was our main interest in this work. 

It should also be noted that the computational method adopted here does not allow one to make a direct estimation of the \paraH2 molecules exchange frequency, unlike its finite temperature counterpart (Path Integral Monte Carlo). Nevertheless, visual inspection of many-particle configurations generated in the Monte Carlo simulation shows little or no overlap of paths associated to different molecules, which is substantial evidence that many-particle permutations are absent in this system. This is consistent with the high degree of localization that molecules experience.

\section{CONCLUSIONS}
\label{conclusions}

Using a numerically exact ground state Quantum Monte Carlo method, we studied  \paraH2 adsorption to krypton pre-plated graphite.   We performed calculations based on a simple model, in which graphite corrugation is ignored, the Kr atoms in the spacer layer are assumed static and point-like, and \paraH2-substrate interactions are given by Lennard-Jones type potentials. 

We find that there are two
stable phases of \paraH2, both solid; one is a monolayer commensurate with the Kr layer, while the other is an incommensurate 
monolayer, compressible within a small range of coverages.  Quantum exchanges of hydrogen molecules are suppressed in this system.   This is similar to what is seen for hydrogen on bare graphite.\cite{nho03} Altogether, this study has provided no evidence of a thermodynamically stable liquid phase of \paraH2 at $T=$0 on the substrate considered here; it is unlikely, based on comparisons with \paraH2 on lithium from Ref. \onlinecite{boninsegni04_li} and 2D \paraH2 from Ref. \onlinecite{boninsegni04b}, that a liquid would appear, in our model, at temperatures as low as 1.5 K.

There are obviously several sources of uncertainty in this calculation which need to be discussed.  The potentials used to describe the interactions between the \paraH2 and the substrate are very rough; this does not seem too important an issue, as far as the interaction of \paraH2 molecules with graphite is concerned, given the relatively large average distance at which molecules sit. On the other hand, a more realistic interaction potential between \paraH2 and krypton may conceivably alter the basic energetics shown here.  Despite these issues, and other simplifications, it does not seem likely (to us) that the basic structural information will change  dramatically.  It is also unlikely that the basic physics will change much at temperatures as low as $\sim$ 1.5 K (namely those reached by the experiment of Ref. \onlinecite{wiechert04}), though finite temperature calculations are planned to address this concern, as are simulations of larger systems.
 
Thus, our preliminary conclusion is that our calculation appears to yield results in disagreement with the experimental findings of Ref. \onlinecite{wiechert04}, reporting a low temperature liquid phase for the heavier isotope of \paraH2 (D$_2$), which should display an even stronger tendency to crystallize than \paraH2, on account of its heavier mass.
\section*{Acknowledgments}
This work was supported in part by the Petroleum Research Fund of the American Chemical Society under research grant 36658-AC5, by the Natural Sciences and Engineering Research council of Canada (NSERC) under research grant G121210893, and by an NSERC PGSB scholarship.  Useful discussions with Milton W. Cole are gratefully acknowledged.


\begin{thebibliography}{9}

\bibitem{ginzburg72}
V. L. Ginzburg and A. A. Sobyanin,  {\it JETP Letters} {\bf 15}, 242 (1972).
\bibitem{bretz81}
M. Bretz and A. L. Thomson, {\it Phys. Rev. B} {\bf 24}, 467 (1981).
\bibitem{maris86}
G. M. Seidel, H. J. Maris, F. I. B. Williams and J. G. Cardon,
{\it Phys. Rev. Lett.} {\bf 56}, 2380 (1986).
\bibitem{maris87}
H. J. Maris, G. M. Seidel and F. I. B. Williams,
{\it Phys. Rev. B} {\bf 36}, 6799 (1987).
\bibitem{schindler96}                                  
M. Schindler, A. Dertinger, Y. Kondo and F.  Pobell,   
{\it Phys. Rev. B} {\bf 53}, 11451 (1996).       
 \bibitem{boninsegni04b}
M. Boninsegni, {\it Phys. Rev. B} {\bf 70}, 193411 (2004).
 \bibitem{boninsegni99}
 M. Boninsegni, M. W. Cole and F. Toigo, {\it Phys. Rev. Lett.} {\bf 83}, 2002 (1999).  
\bibitem{boninsegni04_li}
M. Boninsegni, {\it Phys. Rev. B} {\bf 70}, 125405 (2004).
\bibitem{nielsen80}
M. Nielsen, J. P. McTague and L. Passell in {\it Phase Transitions in Surface Films}, edited by J. Dash and J. Ruvalds (Plenum, New York, 1980).
\bibitem{lauter90}
H. J. Lauter, H. Godfrin, V. L. P. Frank and P. Leiderer in {\it Phase Transitions in Surface Films 2}, edited by H. Taub, G. Torzo, H. J. Lauter and S. C. Fain Jr. (Plenum, New York, 1990).
\bibitem{wiechert91}
H. Wiechert in {\it Excitations in Two-Dimensional and Three-Dimensional Quantum Fluids}, edited by A. F. G. Wyatt and H. J. Lauter (Plenum, New York, 1991).
\bibitem{vilches92}
F. C. Liu, Y. M. Liu and O. E. Vilches, {\it J. Low Temp. Phys.} {\bf 89}, 649 (1992).
\bibitem{wiechert04}
H. Wiechert, K. D. Kortmann and N. St\"usser, {\it Phys. Rev. B} {\bf 70}, 125410 (2004).
\bibitem{wang03}
W.F. Wang, {\it JQSRT} {\bf 76}, 23 (2003).
\bibitem{shi03}
W. Shi, J. K. Johnson and M. W. Cole, {\it Phys. Rev. B} {\bf 68}, 125401 (2003).
\bibitem{steele}
W. A. Steele, {\it Surf. Sci.} {\bf 36}, 317 (1973).
\bibitem{gooding}
R. J. Gooding, B. Joos and B. Bergersen, {\it Phys. Rev. B} {\bf 27}, 7669 (1983).
\bibitem{nho02}
K. Nho and E. Manousakis, {\it Phys. Rev. B} {\bf 65}, 115409 (2002).
\bibitem{gatica04}
S. M. Gatica, J. K. Johnson, X. C. Zhao, and M. W. Cole, {\it J. Phys. Chem. B} {\bf 108}, 11704 (2004).
\bibitem{levesque02}
D. Levesque, A. Gicquel, F. L. Darkrim, and S. B. Kayiran, {\it J. Phys. CM} {\bf 14}, 9285 (2002).
\bibitem{Silvera1}
I. F. Silvera and V. V. Goldman, {\it J. Chem. Phys.} {\bf 69}, 4209 (1978).
\bibitem{johnson96} Q. Wang, J. K. Johnson and J. Q. Broughton, {\it Mol. Phys.} {\bf 89}, 1105 (1996).
\bibitem{operetto}
F. Operetto and F. Pederiva, {\it Phys. Rev. B} {\bf 69}, 024203 (2004).
\bibitem{wang97}
Q. Wang and J.K. Johnson, {\it Mol. Phys.} {\bf 95}, 299 (1998).
\bibitem{macgowan86}
D. MacGowan and Denis J. Evans, {\it Phys. Rev. A} {bf 34}, 2133 (1986).
\bibitem{Ceperley1}
D.M. Ceperley, {\it Rev. Mod. Phys.} {\bf 67}, 279 (1995).
\bibitem{Sarsa1}
A. Sarsa, K. E. Schmidt and W. R. Magro, {\it J. Chem. Phys.} {\bf 113}, 1366 (2000).
\bibitem{boninsegni04a}
M. Boninsegni and L. Szybisz, {\it Phys. Rev. B} {\bf 70}, 024512 (2004).
\bibitem{nho03}
K. Nho and E. Manousakis, {\it Phys. Rev. B} {\bf 67}, 195411 (2003).
\end{thebibliography}
\end{document}